\documentclass[aps,floats,prb,showpacs,twocolumn]{revtex4-1}

\usepackage{graphicx}
\usepackage{dcolumn}
\usepackage{times}
\usepackage{amsfonts,amsmath,amssymb,mathrsfs}
\usepackage{setspace,epstopdf}
\usepackage{epsfig,subfigure}
\usepackage{latexsym}
\usepackage{bbm}
\usepackage{float}
\usepackage{flafter}
\usepackage{bm}
\usepackage{hyperref}
\usepackage{color}
\usepackage{ulem}

\begin{document}

\title{Spin and charge modulations in a single hole doped Hubbard ladder -- verification with optical lattice experiments }

\author{Zheng Zhu$^{1}$, Zheng-Yu Weng$^{1}$, Tin-Lun Ho$^{1,2 *}$ }
\affiliation{$^{1}$Institute for Advanced Study, Tsinghua University, Beijing, 100084, China\\
$^{2}$ Physics Department, The Ohio State University, Columbus, Ohio 43210, USA}

\date{\today}

\begin{abstract}
We show that pronounced modulations in  spin and charge densities can be induced by the insertion of a single hole in an otherwise half-filled 2-leg Hubbard ladder.
Accompanied with these modulations is a loosely bound structure of the doped charge with a spin-1/2, in contrast to the tightly bound case where such modulations are absent.
These behaviors are caused by the interference of the Berry phases associated a string of flipped spins (or ``phase strings") left behind as a hole
travels through a spin bath with a short-range anti-ferromagnetic order. The key role of the phase strings is also reflected in how the system respond to increasing spin polarization, increasing the on-site repulsion, addition of a second hole, and increasing asymmetry between intra- and inter-chain hopping.  Remarkably, all these properties persist down to ladders as short as $\sim 10$ sites.  They can therefore be studied in cold atom experiments using the recently developed fermion microscope.
\end{abstract}


\maketitle

The Hubbard model is a prime example of strongly correlated system. It has a deceptively simple appearance  -- a system of spin-1/2 fermions in a tight binding lattice with onsite repulsion $U$. Yet despite decades of studies, the problem remains unsolved  except in the one-dimensional case. Of particular interest is the 2D Hubbard model, for it is believed that it captures the key physics of high  T$_c$  superconductivity. At half-filling, the ground state of a 2D Hubbard model with strong repulsion is an anti-ferromagnet. There is the expectation that the ground state will become a d-wave superfluid when sufficiently many holes are added. The nature of the ground state as the system is doped away from half filling has been the central question.

In solid state experiments, it is difficult to change the density of electrons continuously, nor is it possible to remove completely the disordered effects that entangle with strong correlation. As a result,  comparison between theory and experiment is not straightforward at times. On the other hand, Hubbard models can now be engineered in cold atom experiments, with easy control of density and interaction\cite{Hubbard,ETH1,Bloch1,ETH2,Bloch2,Hulet}. In principle, one can obtain the solution of the Hubbard model by quantum simulation, i.e. finding the nature of the ground state directly from experiments. Unfortunately, current experiments have not reached temperatures low enough to study strongly correlated effects, due to the heating caused by spontaneous emission.  On the other hand, heating effects can be reduced in small samples, as the low energy excitations in bulk are gapped out by reduced sample size. In addition, the extraordinary development of atom microscope allows one to  image specific atomic species with  single site
resolution\cite{Harvardatom1,Harvardatom2,Blochatom1,Blochatom2,MITatom} making quantum simulations with small systems a powerful way to explore strong correlation effects.
\begin{figure}[tbp]
\centerline{\includegraphics[height=2.65in,width=3in] {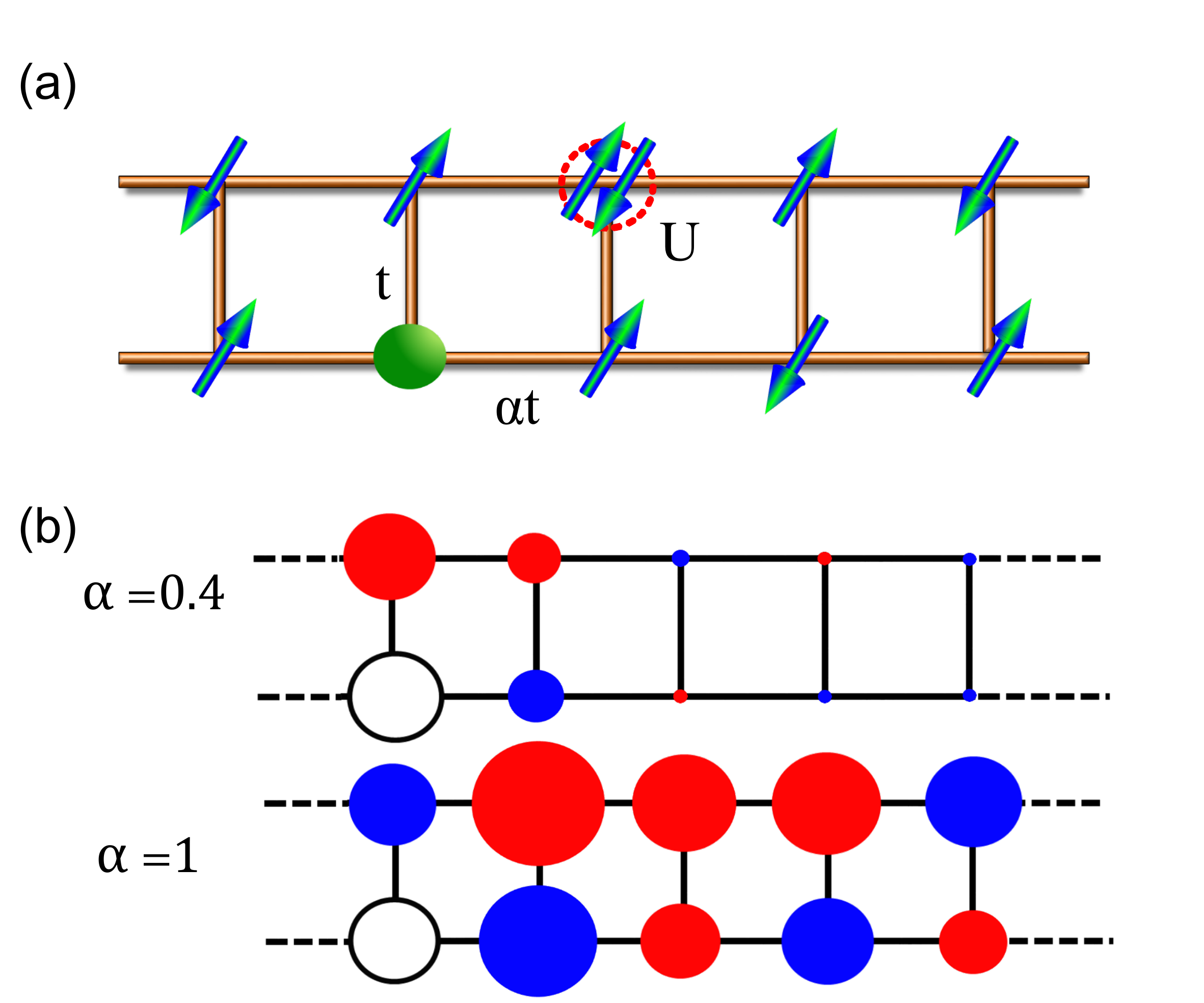}}
\caption{(Color online) (a) The structure and parameters of a 2-leg Hubbard square ladder. Here,  $U$ represents the onsite Coulomb repulsion, $t$ ($\alpha t$) describes the inter-chain (intra-chain) hopping. (b) With one hole (open circle) created by removing a down-spin away from the half-filling spin ladder, a total spin $S_z=1/2$ is found in a spin gapped background in the large $U/t$ (Mott) regime.  Such a total $S_z=1/2$ is found to distribute around the hole with an alternative distribution of up (red circle) and down (blue circle) spins (with the magnitude represented by the size of each full circle). At $\alpha=0.4$, the doped charge is tightly bound with a spin-1/2; in contrast, the hole and the spin-1/2 are loosely bound at $\alpha=1$, where the phase string becomes unscreened as previously discussed in the context of the $t$-$J$ ladder \cite{ZZ2014qp} (see text). }
 \label{Fig:lattice}
\end{figure}
\begin{figure*}[tbp]
\centerline{
    \includegraphics[height=2.0in,width=7.0in] {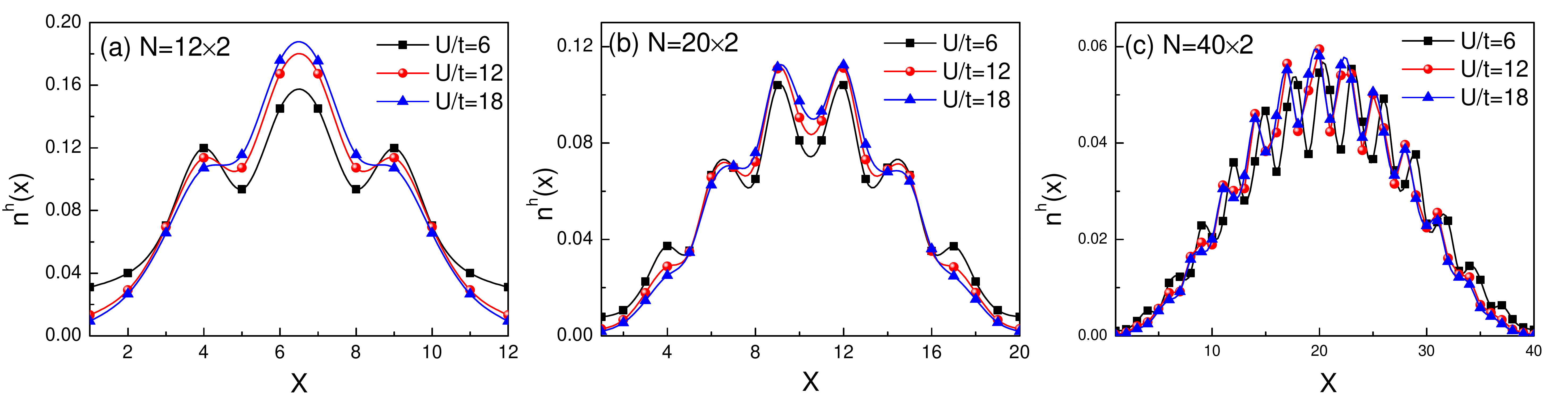}
    }
\caption{(Color online) The real-space charge density distribution along the chain direction of a 2-leg Hubbard ladder doped by one hole at various $U/t$'s. Generically a charge modulation is exhibited in all the sample sizes from $N=12\times2$ (a), $N=20\times2$ (b), and $N=40\times2$ (c). Here the anisotropic parameter is fixed at $\alpha=1$.   } \label{Fig:ChargeModulation}
\end{figure*}

In this paper, we point out some  unusual  properties of a 2-leg Hubbard ladder that reflect
the underlying mechanism  controlling the motion of charge and spin; and explain why quantum simulation with small cold atom systems is a powerful tool for exploration of strong correlation effects. In particular, we  show that :

\noindent
(i) ``Spin and charge modulations": Pronounced modulations in the spin and particle densities can be triggered by the introduction of a single hole into a half filled system. These modulations persist as the length of the ladder is reduced. They remain significant for ladders as short as containing eight sites along the chain direction.

\noindent (ii) ``Spin-charge separation" : Accompanied with the appearance of these modulations is
a loosely bound or composite structure of the charge and spin associated with the doped hole. In fact, the modulations disappear once the spin and charge become tightly bound in an asymmetric limit to be detailed in the paper.

\noindent (iii) ``Phase string effects": These phenomena are caused by the interference of the Berry phases associated with the strings left behind by the hole (or phase strings) as it moves through the spin bath of the half-filled system\cite{phasestring}.  We  point out a number of experimental methods to turn off the phase string effects. Experimental verification of these phenomena will confirm the key role of the ``phase strings".

Similar density modulations have been found by two of us previously (ZZ and ZW) in a two-leg ladder of the t-J model \cite{ZZ2014c}. Our key results (i) to (iii) for the Hubbard model, however, were not contained in Ref. \onlinecite{ZZ2014c}. The fact that  density modulations of bulk sample also occur in small samples means that strongly correlations operate within relatively short range. Such strongly correlations can be  thoroughly explored by performing quantum simulations on small cold atom systems.

\vspace{0,2in}

\begin{figure}[tbp]
\centerline{
    \includegraphics[height=3.5in,width=2.6in] {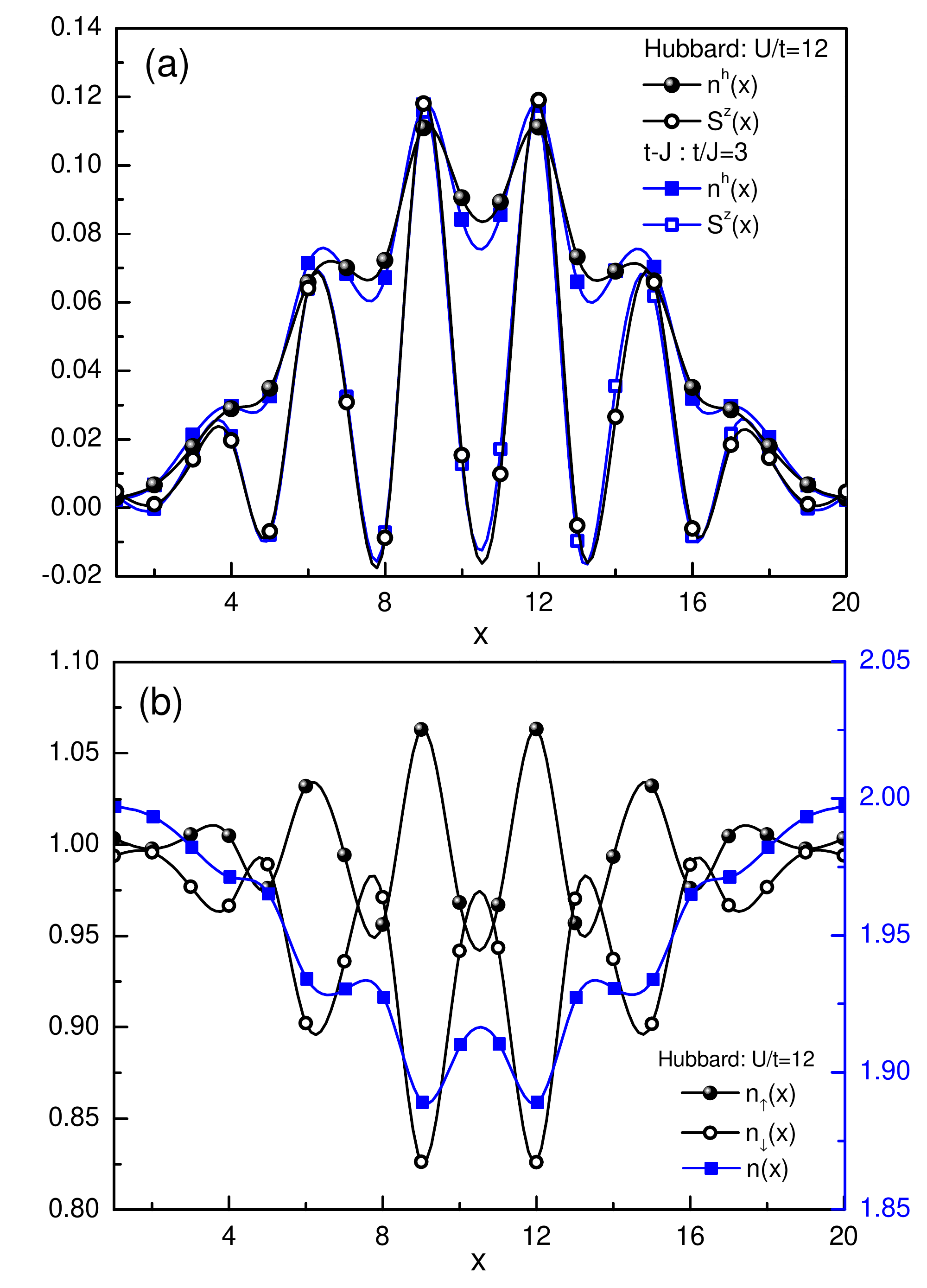}
    }
\caption{(Color online) (a) Novel charge and spin modulations are present in both the Hubbard and $t$-$J$ 2-leg ladder doped by one hole. (b) Then density of both spin component $n_{\uparrow}(x)$ and  $n_{\downarrow}(x)$, as well as the total density $n(x)=n^{}_{\uparrow}(x) + n^{}_{\downarrow}(x)$. Here the 2-leg ladder in the isotropic case of $\alpha=1$ is considered with $N=20\times2$ at $U/t=12$ and $t/J=3$. } \label{Fig:SpinModulation}
\end{figure}

\noindent {\bf A. Charge/spin modulations as fingerprints of strong correlations }

We consider a two-leg Hubbard ladder with size $N=N_x\times2$ as shown in Fig. \ref{Fig:lattice} (a). The hamiltonian
\begin{eqnarray}
H=& - \alpha t \sum_{i=1}^{N_x -1}\sum_{j=1,2} c^{\dagger}_{\sigma}(i+1, j) c^{}_{\sigma} (i, j) + h.c. \nonumber \\
 & - t\sum_{i=1}^{N_x} c^{\dagger}_{\sigma}(i, 1) c^{}_{\sigma} (i, 2) +h.c. \vspace{1.0in}  \nonumber \\
  & + U\sum_{i=1}^{N_x}\sum_{j=1,2} n_{\uparrow}(i,j) n_{\downarrow}(i,j)
\label{H} \end{eqnarray}
where  $\alpha t$ and $t$ are the hopping integrals along and normal to the chain, respectively, and $U$ is the on-site repulsion.
We have studied the case of a single hole injected into the half-filled two-leg ladder using DMRG
with the numerical details similar to those used in Ref. \onlinecite{ZZ2014c} for the $t$-$J$ case. Prior to the insertion of the hole, the density profiles of both spin and charge are simply flat as all the charge and spin fluctuations are gapped in the Mott regime when $U$ is big enough and spins are short-range singlet paired. With the insert of a single hole by taken one spin (e.g., down spin) out, as illustrated in Fig. \ref{Fig:lattice} (b), a total $S_z=1/2$ spin will emerge from the gapped spin background, which distributes around the hole with the averaging on-site up and down spins indicated by red/blue full circles of varying sizes at $\alpha=0.4$, $1$, respectively, determined by the hole-spin correlation function.

Figure \ref{Fig:ChargeModulation} shows that the hole density exhibits a pronounced modulation for the case $\alpha=1$, for ladders from 12 sites  to 40 sites. We find that the charge modulations are present in all finite-size ladders,  only becoming less visible for ladders shorter than 8 sites. Moreover, these modulations remains prominent for smaller $U$ where the system is way from the t-J limit.
Similar modulations in spin density have also been found.  They are not included in Fig.~\ref{Fig:ChargeModulation} to avoid over-crowding.  The spin modulation is shown in Fig.~\ref{Fig:SpinModulation}(a) together with the charge modulation for the case $N=20 \times 2$ and $U/t=12$. In the same figure the spin/charge modulations for the $t$-$J$ model case at $t/J=3$ is also present for comparison. To match quantities that are easily accessible in cold atom experiments, we plot in Fig. ~\ref{Fig:SpinModulation}(b)
the density of both  spin component $n_{\uparrow}(x)$ and  $n_{\downarrow}(x)$, as well as the total density
$n(x)=n^{}_{\uparrow}(x) + n^{}_{\downarrow}(x)$. The period of the modulation is incommensurate, roughly 2 lattice sites.  Near the center of the chain, the density modulation in each spin component can be as large as $15\%$ of their averaged values, which is detectable with atom microscopes. The fact that Fig.~\ref{Fig:SpinModulation} shows a net total spin $S_{z}$ is because we have removed a spin-down fermion. \\

\begin{figure*}[tbp]
\centerline{
    \includegraphics[height=4in,width=6in] {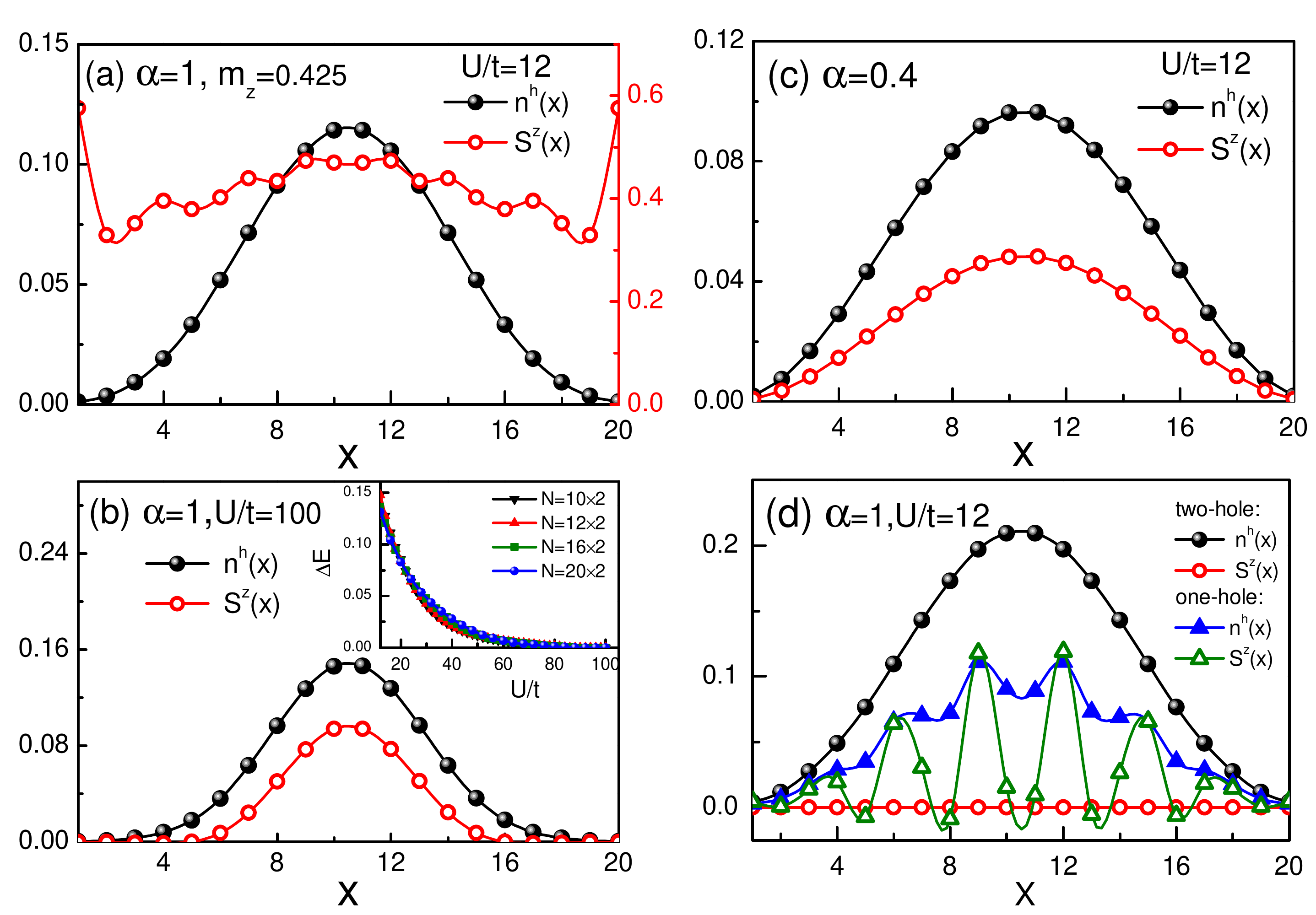}
    }
\caption{(Color online) (a) By polarizing the spin background, the charge modulation can be tuned to vanish. Here $m_z\equiv 2S^z/N$. (b) The charge and spin modulations diminish when the spin singlet-triplet gap $\Delta E$ vanishes at large $U/t$ (the inset); Here the case of $U/t=100$ is shown in the main panel. (c) The charge modulation as the fingerprint of the phase string at $\alpha=1$ is removed in the strong rung case $\alpha=0.4$. (d) The charge modulation seen in the single hole case can be eliminated by the second hole doped into the system, which forms a bound state with the first hole to remove the phase string effect. Here $U/t=12$ and $N=20\times2$.} \label{Fig:NoModulation}
\end{figure*}

\noindent {\bf B. The origin of spin/charge modulations-- phase string effects: }

In the case of the t-J model, it is found that the charge modulation is due to the interference of  ``phase strings". A phase string is hidden in a string of flipped spin left behind as a hole moves in a spin bath with anti-ferromagnetic correlations. Associate with this string is a Berry phase. The total Berry phase accumulated as a hole moves along a closed path $C$ is
\begin{equation}\label{ps}
\tau_C^{ps}\equiv (-1)^{N^{\downarrow}_h[C]}
\end{equation}
where  $N^{\downarrow}_h[c]$ is  the total number of mutual exchanges between the hole and the $\downarrow$-spins as the loop $C$ is traversed. As a hole moves from  point $a$ to point $b$ and back,  different loops  connecting $a$ and $b$ contain different phase string factors $\tau_C^{ps}$, leading to an oscillator in spatial density \cite{ZZ2014c}.

However, for Hubbard model with reduced $U$, the general sign structure is different from the $t$-$J$ model, as holes and doublons can be created through quantum fluctuation. The precise sign structure for the Hubbard model has recently worked out rigorously.\cite{LZhang2014}  Specifically, the partition function of the Hubbard model can be  expressed as\cite{LZhang2014}
\begin{equation}
{\cal Z}=\sum_C \tau_C {\cal W}[C],
\label{Z} \end{equation}
where $C$ denotes the set of closed paths of all particles with a positive weight ${\cal W}[C]>0$.  $\tau_C$  is the sign function
\begin{equation}
\tau_C=(-1)^{N^{\downarrow}_h[C]}(-1)^{N^{\downarrow}_d[C]}(-1)^{N^{ex}_h[C]}(-1)^{N^{ex}_d[C]},
\label{tauC} \end{equation}
where  $N^{\downarrow}_h[C]$ ($N^{\downarrow}_d[C]$) denotes the total number of mutual exchanges between the $\downarrow$-spins and the holons or empty sites (doublons or double-occupied sites) in a given closed path $C$; and $N^{ex}_h[C]$ ($N^{ex}_d[C]$) denotes the total number of exchanges between the holons (doublons).

For large-$U$ and at half-filling, creating a pair of holon and a doublon costs a large energy $U$, and consequently holons and doublons must be created and annihilated in tightly bound pairs so that the factor $(-1)^{N^{ex}_h[C]}(-1)^{N^{ex}_d[C]}$ become +1 for most loops. (Their virtual excitations result in a super-exchange coupling $J=4t^2/U$ between the nearest-neighboring spins.)  In this limit, one has $\tau_C\rightarrow +1$  as in the half-filled $t$-$J$ (i.e., Heisenberg) model. However, with the inserting of a single hole, the phase factor $\tau_C $ in Eq.(\ref{tauC}) reduces to that of
the t-J model Eq.(\ref{ps}), $\tau_C \rightarrow \tau^{ps}_C=  (-1)^{N^{\downarrow}_h[C]} $. The other three factors in Eq.(\ref{tauC}) are absent because there is only one hole and there are no doublons in the large $U$ limit.

\vspace{0.2in}

In the case of the t-J model, one can show mathematically that  the phase strings are the cause of spin and charge modulation by considering an alternate model (referred to as "$\sigma\cdot t$-$J$ model") which augments the hopping of particle with a phase factor that cancels the Berry's phase Eq.(\ref{ps}). For this model, there are no charge modulations with the doping of a single hole \cite{ZZ2014c}.
In the Hubbard case, we are interested in the Mott limit which gives rise to an anti-ferromagnetic insulating ground state in half-filled case, but $U/t$ is still not large enough to reach the $t$-$J$ regime. In this regime, the factor $(-1)^{N^{ex}_h[c]}(-1)^{N^{ex}_d[c]}$ is +1 for most loops. This is because the doublons and holes generated by quantum fluctuations typically form tightly bound pairs in the Mott regime as discussed before. One can then define an analogous ``
$\sigma$-Hubbard model"  to remove the whole $(-1)^{N^{\downarrow}_h[c]}(-1)^{N^{\downarrow}_d[c]} $ by adding a spin-dependent sign $\sigma$ to the hopping term involving exchanges between a single-occupied site (spin) with either a holon or doublon though a projection operator\cite{hongchen jiang}.  The phase string effect is therefore removed completely in the $\sigma$-Hubbard model. Consequently, the charge/spin modulations are indeed absent upon addition of a hole, which has been recently shown by S. X Liu and H.C. Jiang \cite{hongchen jiang} using DMRG, in sharp contrast to the present Hubbard model results shown here, e.g., in Fig.~\ref{Fig:SpinModulation}.

\vspace{0.2in}

\noindent {\bf C. Other ways to remove spin/charge modulations:}

We conclude by pointing a number of physical effects that shows
the phase string mechanism to be the origin for the spin and charge modulations. The idea is to find ways  to diminish the phase string effects, and verify that the spin and charge modulations will disappear in the process.

{\em I. Spin polarization:}
Since the phase string effects are due to the motion of holes in an anti-ferrmagnetic spin background,
they can be manipulated by tuning the spin correlation of the background. This may be achieved by increasing spin polarization of the system.
Indeed the charge modulation gets continuously weakened with the increase of total spin $S_{z}$ as shown in Fig. ~\ref{Fig:NoModulation} (a).

Note that the single hole ground state corresponds to $S_{z}=1/2$, with the magnetization $m_z\equiv 2S_{z}/N= 0.025$ (with $N=20\times 2$). The corresponding charge modulation [cf.  Fig. ~\ref{Fig:SpinModulation}] eventually disappears as $S^z$ is increased to, say, $17/2$ or $m_z =0. 425$ as shown in Fig. ~\ref{Fig:NoModulation} (a).
This disappearance of the charge modulation as the number of down spins are reduced can again be traced back to the phase string effects.  Note that the corresponding spin modulation is also diminished in Fig. ~\ref{Fig:NoModulation} (a).

\noindent{\em II. Large $U/t$:}
At $U/t=12$, the spin background is spin singlet with an extra 1/2-spin loosely bound to the hole [cf. Fig.~1(b)], while at $U/t=\infty$, the ground state becomes spin fully polarized known as the Nagaoka state\cite{Nagaoka}. Then one can expect that in a sufficiently large but finite $U/t$, the spins surrounding the doped hole may still remain polarized, whereas the spin background tends to become singlet, which may be called a Nagaoka polaron state with a finite total spin $S_z>1/2$.
In Fig. ~\ref{Fig:NoModulation} (b), the charge/spin modulations are shown to disappear in a Nagaoka polaron state at $U/t=100$, where the spin singlet-triplet gap [$\Delta E\equiv E_0(S_z=3/2)-E_0(S_z=1/2)$] vanishes as shown in the inset. Thus if one tunes $U/t$ experimentally, the spin and charge modulations can also get removed beyond a critical ratio.

\noindent {\em III. Large hopping asymmetry:}
Spin and charge modulations can also be removed by increasing the asymmetry in the hopping integral. In the limit of $\alpha \gg 1$, the ladder reduces to two 1D chains. Since phase string
interference is absent as there are no close loops that enclose nonzero areas in 1D, there are no density  spin and charge modulations in one hole case. In the opposite limit where $\alpha \ll 1$, the half-filled case corresponds to a singlet pair on every two sites connected vertically $c_{\uparrow}^{\dagger}(i,1)c_{\downarrow}^{\dagger}(i,2)- c_{\downarrow}^{\dagger}(i,1)c_{\uparrow}^{\dagger}(i,2)$. The removal of one fermion will create a localized spin mainly on a vertical rung as
indicated in Fig. \ref{Fig:lattice} (c) for $\alpha =0.4$ in the case of $U/t=12$. Figure ~\ref{Fig:NoModulation} (c) shows that the charge/spin modulations indeed disappear for $\alpha =0.4$, which has been also seen in the $t$-$J$ model case \cite{ZZ2014c}. From the view point of the structure of the partition function in Eq.(\ref{Z}), changing $\alpha$ will change the weight ${\cal W}[C]$, making them only significant if the loops are along the chain ($\alpha\gg 1$) limit or to a single rung ($\alpha\ll1$).
Physically, these two limits corresponds to the total spin-charge separation (1D) and the tightly bound spin-charge inside a quasiparticle [cf. Fig.~1(b)].
In either case, the loop interference of phase string factors become trivial and will not cause density and spin modulations.

\noindent {\em IV. Adding another hole:}
Figure ~\ref{Fig:NoModulation} (d) shows that the charge distribution for the two-hole-doped Hubbard ladder (at $U/t=12$)  is smooth, in sharp contrast to the modulation found in the single-hole case.
This looks as if  the phase string effects of each of these two holes cancel each other, which will be impossible unless the two holes are bound together (but distributed all over the sample). For a bound pair,  the non-trivial sign of the fluctuating phase string associated with each of these two holes are the same.  The total phase string factors then becomes trivial (i.e. +1),  leading to a smooth density profile.  This cancellation of the phase string factor has also been observed previously for the $t-J$ model\cite{ZZ2014}. We have verified here that the cancellation persists away from the $t-J$ limit.  The disappearance of the spin and charge modulations therefore reflects the binding of holes through phase strings, which is a non-BCS pairing force.

{\em Concluding Remarks:}  The mechanism that controls the nature of the ground state of the Hubbard model in the Mott regime has been a question of central interests. Our studies of the 2-leg Hubbard ladder show that the phase string effect is central to the organization of spin and charge degrees of freedom. It is truly timely that  cold experiments have advanced to the point the physical conditions discussed here can be engineered, and all the phenomena discussed here can be studied with the newly developed fermion microscope.
Not only will these experiments verify the dramatic spin and density modulations triggered by a single doped hole, but also verify the key role of phase strings. It is conceivable that other  strongly correlated effects will persist down to a small sample.  Quantum simulation of small cold atom systems can be a new powerful way to study strongly correlated phenomena of bulk systems.

{\em Acknowledgement:} Stimulating discussions with Rong-Qiang He, Hongchen Jiang, and Shenxiu Liu are acknowledged. This work is supported by the NBRC grant 2015CB921000  (ZW); the NSF grant DMR-1309615, the MURI grant FP054294-D, and the NASA grant 1501430 awarded to TLH.

\vspace{0.2in}

* jasontlho@gmail.com

\end{document}